**Edge energy fluctuation: A unified theory of friction scaling laws in twisted layered material interfaces**

Ze Liu [a,b *]

[a] *Department of Engineering Mechanics, School of Civil Engineering, Wuhan University, Wuhan, Hubei 430072, China*

[b] *State Key Laboratory of Water Resources & Hydropower Engineering Science, Wuhan University, Wuhan, Hubei 430072, China*

* Corresponding authors at: Wuhan University, Department of Engineering Mechanics, Wuhan, Hubei 430072, China

E-mail address: ze.liu@whu.edu.cn (Z. Liu).

**Abstract:** Interlayer twisting in van der Waals (vdW) layered materials generates moiré superlattices, unlocking unique opportunities for engineering their electronic and mechanical functionalities. Here, building on two key insights – the invariance of the potential energy of complete moiré tiles during sliding, and the symmetry-breaking effect induced by the slider edge – we propose the concept of edge energy fluctuation (EEF) and demonstrate that the EEF governs the sliding energy barrier of finite-sized sliders. Interestingly, we found that even at a fixed twist angle and sliding direction, significant frictional anisotropy emerges when the slider edge is cut along different directions, resulting in distinct friction scaling laws. Furthermore, we present a friction engineering paradigm for twisted layered material interfaces through the deliberate introduction of hierarchical edges, offering a promising route for friction control in superlubric devices.

When two identical or similar two-dimensional (2D) lattices are stacked with a relative twist angle, the resulting moiré superlattice disrupts interlayer registry, causing sliding lateral forces from individual atoms to largely cancel [1,2]. This phenomenon – known as structural superlubricity, first realized in twisted nanoscale graphite contacts [3] – has since been extended to large scales [4-7] and heterogeneous interfaces [8-10], becoming a transformative paradigm for energy-efficient micro- and nano-scale mechanical systems [1,11-14].

However, extensive experimental and computational studies have revealed that friction can still occur at a twisted interface of finite size, following nontrivial scaling laws with respect to contact area ($A$) and twist angle ($\theta$) that deviate significantly from classical Amontons' law. Experimental studies on 2D contact interfaces have reported a broad range of scaling exponents $m$ in the friction scaling laws of $F \propto A^m$, spanning from 0 to 0.5 [15-19]. Recent theoretical work has clarified this scatter by considering the interplay of size and shape effects, as well as the relative orientation between the slider edges and moiré superlattices [20-22]. Notably, a unique double periodicity of static friction with contact size and the absence of size scaling for twisted incommensurate polygonal slides have been discovered [21].

Nevertheless, a critical gap in current understanding persists: most computational or theoretical models assume periodic boundary conditions and ideal shapes, whereas realistic sliders usually have finite and irregular geometric shapes. In this work, we address this gap by developing a unified theoretical framework for interfacial sliding in finite-sized twisted vdW contacts, founded on the proposed concept of edge energy fluctuation (EEF). We demonstrate that the EEF governs the sliding energy barrier and derive a general edge integral expression for predicting the friction scaling laws of twisted layered materials with arbitrary shapes.

Figure 1a shows a typical model system consisting of a rigid circular graphene slider deposited on a fixed graphene substrate (Fig. 1a). Recent studies have revealed that the incomplete moiré tiles near the slider edge (denoted by the red-colored region in Fig. 1a), i.e., moiré edge, dominate the friction scaling of vdW layered materials [20,21,23,24]. Taking a circular tBLG with a twist angle of 2° for example, the size scaling of the energy barrier for different contact areas were first calculated through MD simulations (black squares) (Fig. 1b) [20], then the contribution of the energy barrier from the internal region (green triangles) and moiré edge region (red circles) was separated, it is observed that the energy barrier is almost entirely contributed from the moiré edge (Fig. 1b).

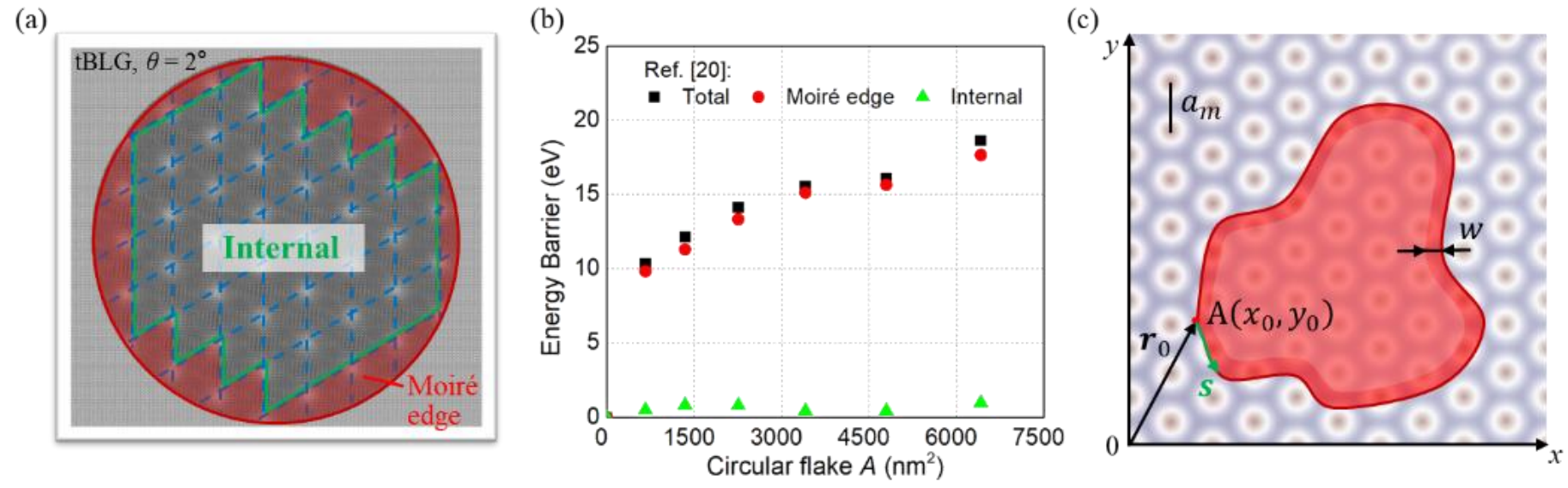

Fig. 1. Definition of edge energy. (a) Illustration of the moiré edge, referring to the incomplete moiré tiles resulting from symmetry breaking at the edge of a circular slider with a twist angle of 2°. (b) Size scaling of the energy barrier for circular tBLGs with a twist angle of 2° [20]. (c) The edge energy of a slider with arbitrary shape is defined in terms of the interlayer potential energy integrated over an edge strip of width $w$.

Physically, for a given twist angle, the energy of any atom within each of moiré superlattices falls within the range of $[\epsilon_{AB}, \epsilon_{AA}]$, where $\epsilon_{AA}$ and $\epsilon_{AB}$ correspond to the atom being above the vertex and above the center of the six membered ring of the bottom graphene, respectively. Therefore, by choosing an energy range $(\epsilon_i, \epsilon_i + d\epsilon_i]$, where $\epsilon_i \in [\epsilon_{AB}, \epsilon_{AA}]$, within which there is only one atom in any of the moiré superlattices, then one-to-one correspondence can be established between atoms locating in different moiré superlattices. In other words, the sum of interlayer potential energies of all atoms in a complete moiré tile will be independent on location. Whereas for incomplete moiré tiles near edges, they are usually incoherent in phase during sliding. Therefore, we define the edge energy (EE) of a slider in terms of the interlayer potential energy integrated over an edge strip of width $w$ (Fig. 1c), and the edge energy fluctuation (EEF) refers to the variation of the integrated potential energy during sliding, which as will be shown below, governs the sliding energy barrier.

By treating a slider as a continuum surface, the moiré-induced periodic potential experienced by an infinitesimal surface area of the slider reads [25-28],

$$U(x,y) = -\frac{2}{9}U_0\left(2\cos\frac{2\pi x}{\sqrt{3}a_m}\cos\frac{2\pi y}{a_m} + \cos\frac{4\pi x}{\sqrt{3}a_m}\right) \quad (1)$$

where $U_0$ is the amplitude of the potential energy landscape corrugation per unit area. $a_m = \frac{a}{2\sin\theta/2}$ is the period of the moiré superlattices, $a = 2.46$ Å is the period of the hexagonal graphene lattice, and $\theta$ is the twist angle. For simplicity, the size of the moiré edge can be estimated analytically by approximating the edge region as a belt with a width of $w = \lambda a_m$ (Fig. 1c), where $\lambda < 1$ is the proportionality coefficient. Then we can derive a generalized edge integral expression for the edge energy of a slider with arbitrary shape (Fig. 1c) as

$$E_{\text{edge}}(x_0, y_0) = \oint U(x,y) \cdot w ds \quad (2)$$

where point $\mathrm{A}(x_0, y_0)$ is a reference point in the slider surface, and $s$ is the magnitude of the unit tangential vector along the slider edge, and its positive direction is determined by the right-hand rule.

To demonstrate the applicability of the edge energy fluctuation, we take the well-studied circular slider as an example (Fig. 2a). For convenience, we set the center of the circular slider as the reference point $\mathrm{A}(x_0, y_0)$, then the edge energy can be derived according to Eq. (2) (see supplementary materials for the detail),

$$E_{circle}(x_0, y_0) = -\frac{4\pi\lambda}{9}U_0 a_m R(\cos\phi_1 + \cos\phi_2 + \cos\phi_3)J_0\left(\frac{4\pi R}{\sqrt{3}a_m}\right) \quad (3)$$

where $\phi_1 = \frac{2\pi}{\sqrt{3}a_m}x_0 + \frac{2\pi}{a_m}y_0$, $\phi_2 = \frac{2\pi}{\sqrt{3}a_m}x_0 - \frac{2\pi}{a_m}y_0$, $\phi_2 = \frac{4\pi}{\sqrt{3}a_m}x_0$, $J_0(\cdot)$ is the Bessel function of the first kind. Based on Eq. (3), the location-dependent interlayer potential energy can be readily obtained. Figure 2b displays typical results for a graphene circular slider of radius $R$ = 50 nm at a twist angle of 5°. For comparison, the potential energy integrated over the full circular domain is shown in Fig. 2c. Remarkably, the two potential energy landscapes are nearly identical, differing only by a 90° phase shift. This phase difference originates from the fact that radial integration over the circular domain converts the zeroth-order Bessel function of the first kind into its first-order counterpart, which, however, only induces a translational shift of the potential energy landscape (Fig. 2b-c), without changing the sliding energy barrier.

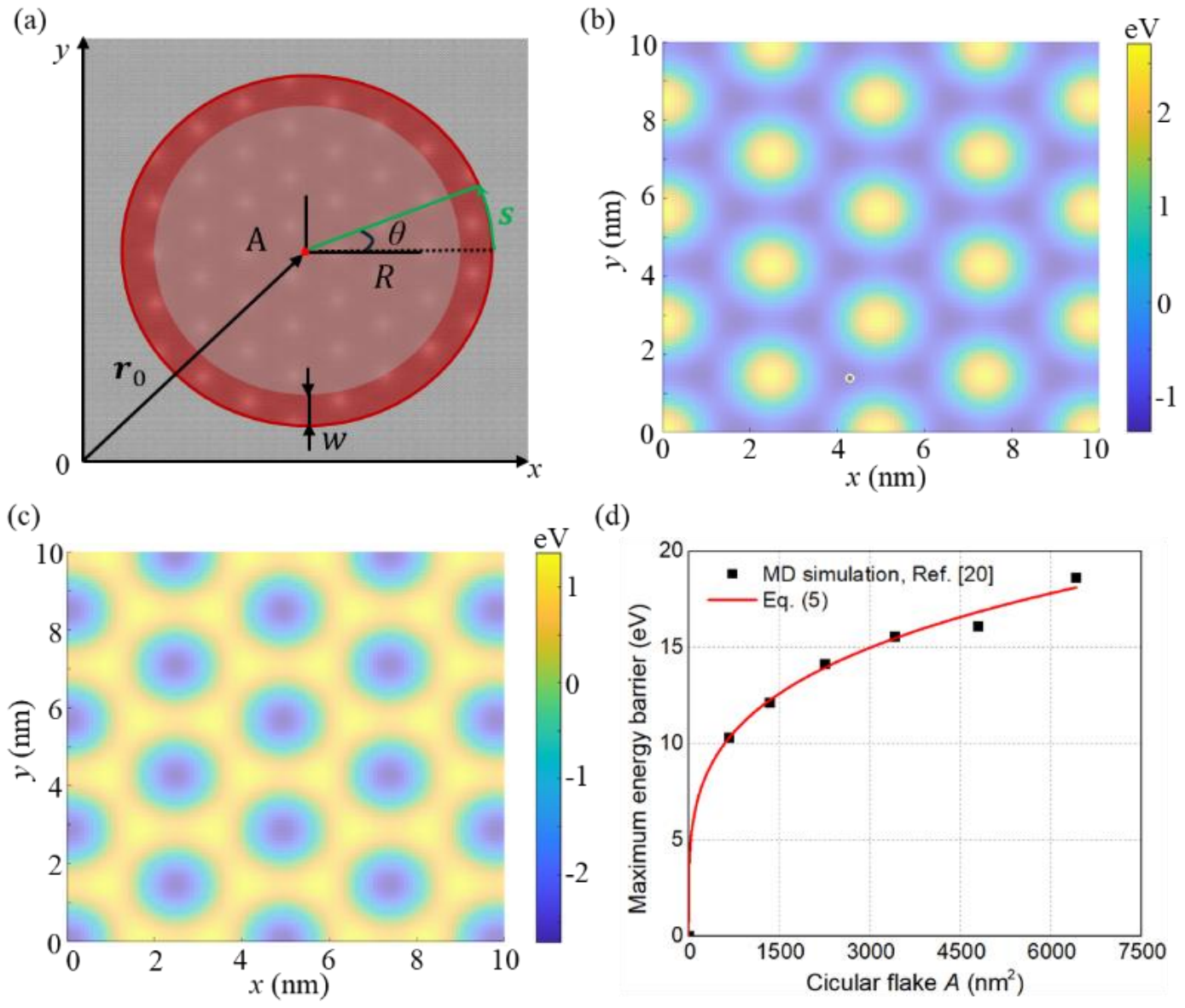

Fig. 2. (a)-(c) Sliding potential energy of a circular graphene slider atop an infinite graphene substrate, where $U_0 = 5.85$ meV/Å$^2$ [21], the radius of the circular slider and the twist angle are 50 nm and 5°, respectively. $\lambda$ is set to 0.191 to exactly reproduce the energy amplitude obtained from the full circular domain integral. (d) Scaling law for a circular slider derived from the edge energy fluctuation theory at a fixed twist angle of 2°. The maximum energy barrier data, obtained from large-scale molecular dynamics simulations, are reproduced from Ref. [20].

According to Eq. (3), the sliding energy barrier can be obtained. Taking, as an illustrative example, a circular tBLG with a twist angle of 2° and assuming the sliding direction coincides with the $x$-axis, the sliding energy barrier is given by

$$\Delta G_{\text{circle}} = \lambda a_m \left| E_{\text{circle}}(0,0) - E_{\text{circle}}\left(\frac{a_m}{\sqrt{3}}, 0\right) \right| = \sqrt[4]{12}\lambda U_0 a_m^{3/2} R^{1/2} \left| \cos\left(\frac{4\pi}{\sqrt{3}a_m} R - \frac{\pi}{4}\right) \right| \quad (4)$$

Here, the asymptotic cosine representation of the Bessel function for large arguments is adopted, as the slider size is generally orders of magnitude larger than the moiré superlattice period. Then the maximum sliding energy barrier can be obtained as

$$\Delta G_{\text{circle,max}} = \sqrt[4]{12}\lambda U_0 a_m^{3/2} R^{1/2} \propto \theta^{-3/2} A^{1/4} \quad (5)$$

As anticipated, the scaling law in Eq. (5) with respect to twist angle and contact area is consistent with previously reported results [20,21]. The prefactor $\lambda$ can be obtained by fitting experimental or simulated data. Taking circular tBLGs with a twist angle of 2° as an example, we perform a fit of Eq. (5) to the maximum energy barrier data acquired from large-scale molecular dynamics simulations over a range of contact areas [20], from which we determine $\lambda = 0.132$ (Fig. 2d).

It is well established that friction between two contacting single crystals displays anisotropy, attributable to differences in either the shearing direction or the lattice orientation [29,30]. Here, we reveal that friction anisotropy can also arise from the edge orientation of a slider, even when the twist angle and sliding direction are held fixed. To demonstrate this, we consider a straight slider edge oriented at an angle $\theta$ relative to the $x$-axis (Fig. 3a), which can be parameterized as

$$\begin{pmatrix} x \\ y \end{pmatrix} = \begin{pmatrix} x_0 \\ y_0 \end{pmatrix} + s\begin{pmatrix} \cos\theta \\ \sin\theta \end{pmatrix}, \quad s \in [0, L] \tag{5}$$

Substituting Eq. (5) into Eq. (1) and applying the edge energy integral expression, we obtain

$$E_\theta(x_0, y_0) = \int_0^L U(x,y) w ds = -\frac{2\lambda}{9} U_0 a_m \sum_{j=1}^{3} \int_0^L \cos(\phi_j + k_j s)\, ds \tag{6}$$

where $k_1 = \frac{2\pi}{\sqrt{3}a_m}\cos\theta + \frac{2\pi}{a_m}\sin\theta$, $k_2 = \frac{2\pi}{\sqrt{3}a_m}\cos\theta - \frac{2\pi}{a_m}\sin\theta$, $k_2 = \frac{4\pi}{\sqrt{3}a_m}\cos\theta$. The integral term in Eq. (6) yields

$$\int_0^L \cos(\phi_j + k_j s)\, ds = \begin{cases} L\cos\phi_j, & k_j = 0 \\ \frac{\sin(\phi_j + k_j L) - \sin\phi_j}{k_j}, & k_j \neq 0 \end{cases} \tag{7}$$

Based on Eqs. (6) - (7), the contribution of this edge to the sliding energy barrier of a slider can be obtained. Taking a tBLG slider with a twist angle of 5° as an example, and assuming the sliding direction is along the $x$-axis, the sliding energy barrier reads

$$\begin{aligned} \Delta E_\theta &= \lambda a_m \left| E_\theta(0,0) - E_\theta\left(\frac{a_m}{\sqrt{3}}, 0\right) \right| \\ &= \frac{4\lambda a_m U_0}{9} \left| \sum_{i=0}^{3} \left( \frac{1}{k_i} \sin\left(\frac{k_i L}{2}\right) \left( \cos\left(\phi_i(0,0) + \frac{k_i L}{2}\right) - \cos\left(\phi_i\left(\frac{a_m}{\sqrt{3}}, 0\right) + \frac{k_i L}{2}\right) \right) \right) \right|, \quad k_j \neq 0 \end{aligned} \tag{8}$$

From Eq. (8), we obtain the edge orientation-dependent sliding energy barrier (Fig. 3b), which exhibits remarkable friction anisotropy. This anisotropy possesses sixfold rotational symmetry, reflecting the underlying symmetry of the moiré-induced potential. When the edge is oriented at $\theta = \frac{n\pi}{3} + \frac{\pi}{6}$ ($n = 0, 1, 2$), corresponding to the friction peaks in Fig. 3b, all incomplete moiré tiles intersected by the edge maintain the same phase throughout sliding – termed a “coherent edge" – independent of the sliding direction. Consequently, the sliding energy barrier is proportional to the edge length, leading to a size-dependent friction scaling law, as verified by large-scale molecular dynamic simulations [21]. However, when the edge is oriented along $\theta = \frac{n\pi}{3}$ ($n$ = 0, 1, 2, 3), corresponding to the friction valleys in Fig. 3b, the sliding energy barrier can be reduced to ideally zero when the edge length satisfies $L = \frac{\sqrt{3}}{2} n a_m$ ($n \in \mathbb{N}$). We refer this configuration as an "incoherent edge".

According to previous studies, superlubric sliding of a slider is often plagued by persistent finite friction originating from incomplete moiré tiles at edges [21,31-33]. Our findings indicate that when a slider is engineered with incoherent edges, the friction arising from incomplete moiré tiles can be suppressed to realize an ideal superlubricity state, in good agreement with recent simulations [21,22]. In contrast, for any other edge orientation, the sliding energy barrier exhibits size-independence [Eq. (8)]. We therefore conclude that friction is size-independent for sliders devoid of coherent edges, a conclusion supported by recent observations on twisted incommensurate polygonal sliders [21]. Notably, this frictional anisotropy originates solely from the slider edge orientation, fundamentally differentiating it from previously reported mechanisms – namely, twist angle-dependent friction in the context of structural superlubricity [3,4,34], and sliding direction-dependent friction induced by the lattice anisotropy of the contact interface[30,35,36].

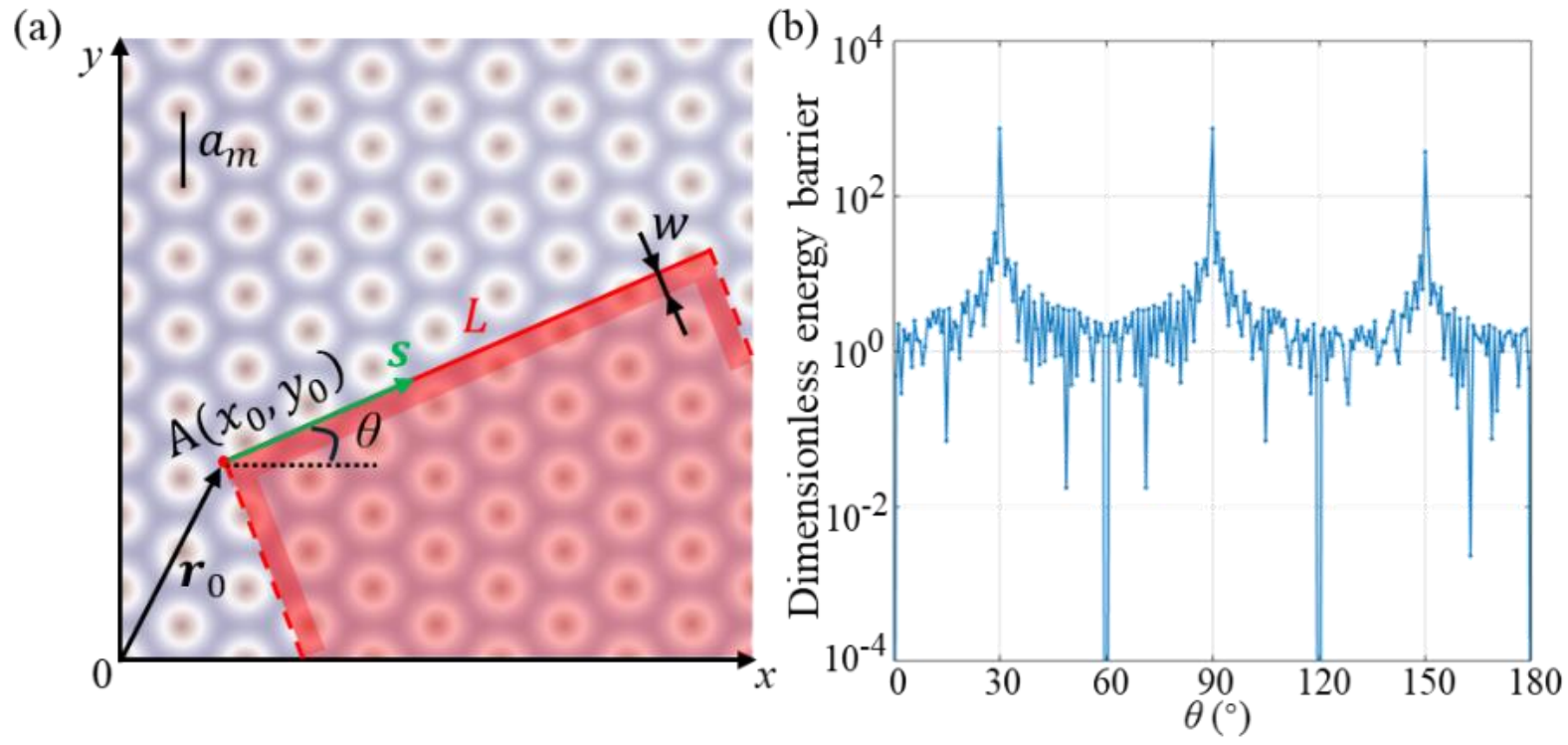

Fig. 3. Friction anisotropy caused by slider edge orientation. (a) Schematic of a slider starting from point $A(x_0, y_0)$ and oriented at an angle $\theta$ to the $x$-axis. (b) Sliding energy barrier as a function of edge orientation, as calculated from Eq. (8) for an edge length of $100\sqrt{3}a_m$ and a twist angle of 5°. The sliding direction is taken to be along the $x$-axis.

Multicontact paradigm provides a potential way for achieving large-scale robust superlubricity [37], exemplified by the incorporation of a high density of nanographite sheets [38,39] or diamond nanoparticles [6] at the contact interface. Within this framework, the edge energy fluctuation theory constitutes an effective theoretical approach for analyzing the frictional behavior of interfaces possessing topological structures. Consider a ring-shaped graphene slider with inner and outer radius of $r$ and $R$, respectively (Fig. S1), the edge energy can be similarly obtained as (see supplementary materials for the detail)

$$E_{\mathrm{ring}}(x_0, y_0) = -\frac{4\pi\lambda}{9} U_0 a_m(\cos\phi_1 + \cos\phi_2 + \cos\phi_3)\left[RJ_0\left(\frac{4\pi R}{\sqrt{3}a_m}\right) - rJ_0\left(\frac{4\pi r}{\sqrt{3}a_m}\right)\right] \quad (9)$$

Assuming the sliding is along the $x$-axis, then the maximum sliding energy barrier gives

$$\Delta G_{ring} = \left|E_{ring}(0,0) - E_{ring}\left(\frac{a_m}{\sqrt{3}}, 0\right)\right|$$
$$= 12^{\frac{1}{4}}\lambda a_m^{\frac{3}{2}} U_0 \left|\sqrt{R}\cos\left(\frac{4\pi R}{\sqrt{3}a_m} - \frac{\pi}{4}\right) - \sqrt{r}\cos\left(\frac{4\pi r}{\sqrt{3}a_m} - \frac{\pi}{4}\right)\right| \quad (10)$$

If $r \ll R$, then Eq. (10) degrades to the scaling law of circular sliders [Eq. (4)]. While if the width of the ring $d = R - r \ll R$, then we have

$$\Delta G_{ring} = \frac{4\sqrt{2}}{\sqrt[4]{3}}\lambda\pi a_m^{\frac{1}{2}} d\sqrt{R}U_0\left|\sin\left(\frac{4\pi r}{\sqrt{3}a_m} - \frac{\pi}{4}\right)\right| \propto a_m^{\frac{1}{2}} d\sqrt{R}U_0 \quad (11)$$

Equation (11) indicates that the scaling law of thin ring sliders with respect to twist angle will become $\theta^{-1/2}$, existing much weaker twist angle-dependence compared to circular sliders ($\theta^{-3/2}$).

Although ultralow friction offers a solution to the persistent issue of device failure caused by friction and wear, it can also present difficulties for sliding devices requiring precise control of motion states. Therefore, the capacity to achieve on-demand regulation of interfacial friction would be highly beneficial for the design and control of superlubric [40,41] and reconfiguration devices [12-14]. The frictional anisotropy induced by the orientation of the slider edge offers a practical and effective strategy for friction control in superlubric devices. As an illustration, by exploiting the symmetry of the moiré-induced potential field, we can harness the properties of a coherent edge. Introducing a hierarchical structure at the edge, typically such as a coherent edge of length $L$ (Fig. 4a) that propagates along other coherent directions in a zigzag pattern, then the edge energy of this hierarchical coherent edge can be correlated with the hierarchical level ($n$) as

$$E_n = -\frac{4(2n+1)}{27}\lambda\pi U_0 a_m L \tag{12}$$

Because the maximum sliding energy barrier is proportional to the edge energy, then we have

$$\overline{\Delta G_{n,\max}} = \frac{2(2n+1)}{3} \tag{13}$$

where the maximum sliding energy barrier is normalized with respect to that of the edge along Path 0 in Fig. 4a. The results demonstrate that, while the friction scaling law is the same as that for a straight edge, the friction magnitude increases linearly with the hierarchical level (Fig. 4b). This result stems from the fact that the potential energy integrated along the line connecting any two adjacent extremum points of the moiré-induced potential field is a constant, $-\frac{2}{9}U_0\lambda a_m^2$, independent of the integration direction. Essentially, this strategy fully utilizes the coherent directions inside the slider through a fractal-inspired design.

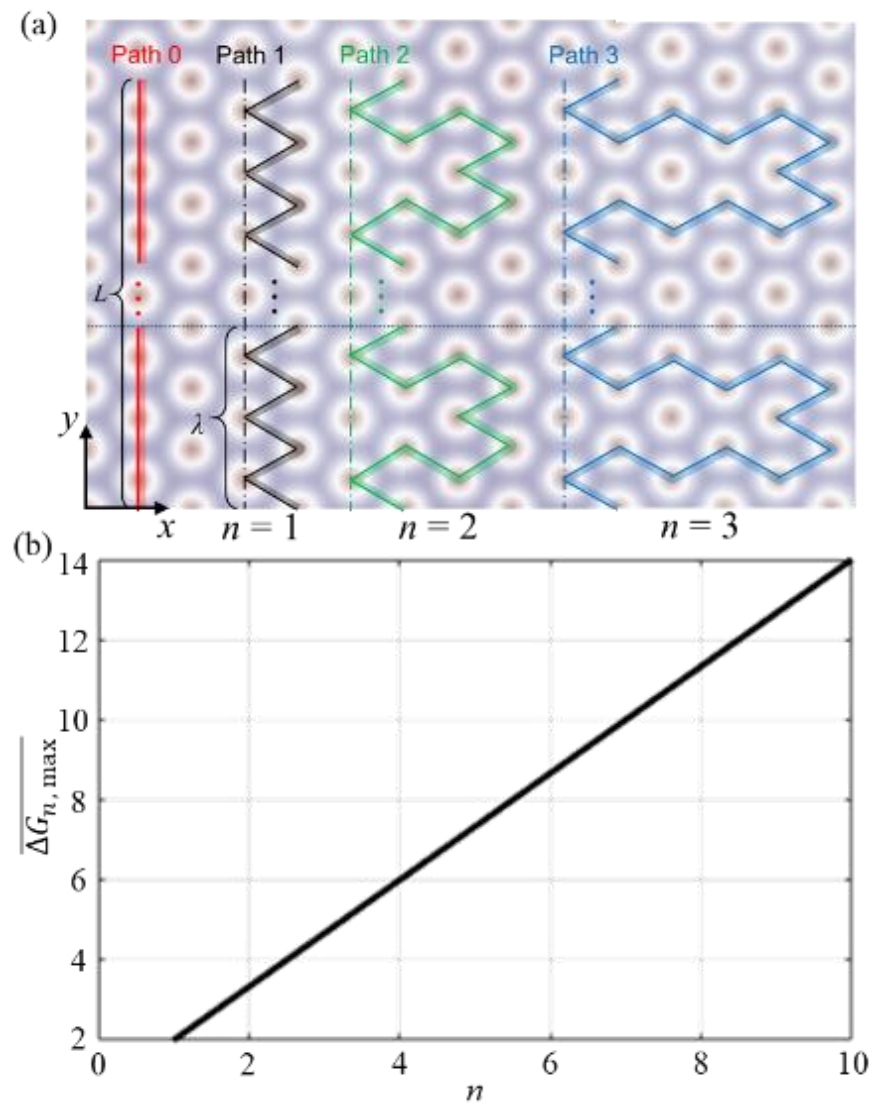

Fig. 4. Friction engineering by introducing hierarchical coherent edges. (a) Sketches of a zigzag-type hierarchical coherent edge. (b) Hierarchical level-dependent maximum sliding energy barrier [Eq. (13)], where the projected length of all edges on the $y$-axis is $L$, and the sliding is along the $x$-axis (a).

The core physics underlying the EEF theory is the sum of the potential energies of all atoms within a complete moiré tile at different positions is approximately a constant, which is guaranteed by the one-to-one positional mapping of atoms between any two complete moiré tiles [42]. Therefore, the sliding energy barrier is governed by the potential energy fluctuation of incomplete moiré tiles at the edge. Such a potential energy fluctuation changes as sliding. As for straight edge, if the potential energy of all incomplete moiré tiles at the edge varies in the same manner during sliding (i.e., the coherent edge), then the friction will scale as the edge length. In contrast, for incoherent edge orientations, previous studies have demonstrated that, with increasing the edge length, the geometries of incomplete moiré tiles on the same or opposite edges can become perfectly complementary. As a result, the maximum sliding energy barrier or static friction force becomes independent of the slider size [21,23,25]. From a mathematical standpoint, Kronecker's approximation theorem [43] guarantees that any straight line must intersect the center of at least one moiré superlattice. Hence, when the opposing edge of the slider likewise passes through a moiré superlattice center, the incomplete moiré tiles along the two edges will exhibit perfect complementary, resulting in a vanishing net contribution to the sliding energy barrier.

This complementary coherence, exhibited by straight edges, ceases to hold for curved edges possessing finite curvature. A typical case is sliders with circular edge (Fig. 2a), the area of incomplete moiré tiles in its edge region can be approximated by $A_{\mathrm{edge}} \sim \pi R a_m$, then the number of incomplete moiré tiles in the edge region reads

$$N_{\mathrm{edge}} \approx \frac{A_{\mathrm{edge}}}{A_{\mathrm{m}}} = \frac{2\pi R}{\sqrt{3} a_m} \tag{14}$$

where $A_{\mathrm{m}} = \frac{\sqrt{3}}{2} a_m^2$ is the area of a complete moiré tile. Since the circular edge intersects the moiré superlattices along a curve, the phase of the potential energy variation $\delta u_i$ ($i = 1,2 \ldots, N_{\mathrm{edge}}$) associated with each incomplete moiré tile during sliding is randomized. Consequently, the sliding potential energy of each incomplete moiré tile, $\delta u_i$, may be considered as a random variable with respect to sliding distance. Then $\sum_{i=1}^{N_{\mathrm{edge}}} \delta u_i$ determines the sliding potential energy change of the circular slider. According to the central limit theorem, the expected value of the amplitude of the sum of $N$ random variables, i.e. the maximum sliding energy barrier, can thus be given as

$$\Delta G_{\max} \sim \sqrt{N_{\text{edge}}} \cdot \langle \delta u_i \rangle \tag{15}$$

where $\langle \delta u_i \rangle \sim A_{\text{m}} U_0 = \frac{\sqrt{3}}{2} U_0 a_m^2$ is the energy fluctuation of each incomplete moiré tile during sliding. Substituting Eq. (14) into Eq. (15) gives

$$\Delta G_{\max} \sim \sqrt[4]{\frac{3\pi^2}{4}} U_0 R^{1/2} a_m^{3/2} \tag{16}$$

At small twist angles, $a_m = \frac{a}{2\sin\frac{\theta}{2}} \approx \frac{a}{\theta}$, then we have

$$\Delta G_{\max} \propto \theta^{-3/2} R^{1/2} \tag{17}$$

which yields a scaling law identical to that obtained from the EEF theory [Eq. (5)] and documented in previous studies [20,21].

To summary, twisted vdW layer material interfaces afford an unprecedented avenue for substantially regulating material properties via interlayer twist, serving as a powerful complement to conventional compositional modification. Our results reveal that, in addition to the twist angle, the slider geometry—most notably its edge orientation—exerts a profound influence on the interfacial sliding behavior of vdW layered materials, yielding qualitatively different friction scaling laws. We elucidate that the unique friction scaling laws find their physical origin in the symmetry breaking of the moiré-scale potential by the slider edge, a mechanism that directly engenders frictional anisotropy even under conditions of fixed interlayer twist angle and sliding direction. Notably, all of the aforementioned findings originate from the EEF framework developed in this study. Recognizing that a fundamental challenge in moiré physics is the artificial design of moiré superlattices for engineering the macroscopic response of twisted vdW layered materials, we expect that the EEF theoretical framework proposed herein will not only help to elucidate the long-standing puzzle concerning friction scaling laws at vdW layered material interfaces, but also offer a predictive design guideline for twistronics [12,44,45] and superlubric devices [11,13,14] requiring controlled friction.

**Acknowledgements** Z. L. would like to acknowledge supports from the National Key Research and Development Program of China (2025YFF0513900), and the National Natural Science Foundation of China (No. 92477138).